\documentclass{ws-ijbc}
\usepackage{graphicx}

\begin{document}

\catchline{}{}{}{}{} 

\markboth{Mois\'{e}s Santill\'{a}n}{Periodic forcing of a 555-IC based electronic oscillator in the strong coupling limit}

\title{Periodic forcing of a 555-IC based electronic oscillator in the strong coupling limit}

\author{Mois\'{e}s Santill\'{a}n}

\address{Unidad Monterrey, Centro de Investigaci\'{o}n y Estudios Avanzados del IPN, V\'{\i}a del Conocimiento 201, Parque PIIT, 66600 Apodaca NL, M\'{E}XICO\\ msantillan@cinvestav.mx}

\maketitle

\begin{history}
\received{(to be inserted by publisher)}
\end{history}

\begin{abstract}
We designed and developed a master-slave electronic oscillatory system (based on the 555-timer IC working in the astable mode), and investigated its dynamic behavior regarding synchronization. For that purpose we measured the rotation numbers corresponding to the phase-locking rhythms achieved in a large set of values of the normalized forcing frequency (NFF) and of the coupling strength between the master and the slave oscillators. In particular we were interested in the system behavior in the strong-coupling limit, because such problem has not been extensively studied from an experimental perspective. Our results indicate that, in such a limit, a degenerate codimension-2 bifurcation point at NFF=2 exists, in which all the phase-locking regions converge. These findings were corroborated by means of a mathematical model developed to that end, as well as by \textit{ad hoc} further experiments.
\end{abstract}


\section{Introduction}
\label{intro}

The phenomenon of synchronization has fascinated scientists since its discovery by Christiaan Huygens circa 1666, but only in the late 20th century was a solid mathematical theory developed that allowed us to deeply understand some of the simplest examples of synchronization \cite{Peskin:1975uq,Mirollo:1990kx}. Despite these and other seminal achievements, the problem in general is still open for research---see for instance \cite{Pikovsky:2001fk}, Chapter 8.

A simple alternative to the problem of two mutually-interacting oscillators is having one of them (the master oscillator) unidirectionally influencing the other (the slave oscillator). This problem has been extensively studied, and actually we know what to expect in the weak coupling limit---see \cite{Pikovsky:2001fk}, Chapter 7. Whenever the master and slave oscillator frequencies are similar and the interaction between them is strong enough, they achieve 1:1 synchronization. When at least one of these conditions is not satisfied, one can obtain either other phase locked rhythms or aperiodic (quasi-periodic) rhythms. In the two-parameter stimulation-frequency---forcing-intensity plane, the former rhythms occur within areas known as Arnol'd tongues, while the latter occur in one-dimensional arcs interlaced between the Arnol'd tongues \cite{Boyland:1986ys}. This behavior has been observed, both experimentally and theoretically, in multiple physical, chemical, and biological oscillators \cite{Boyland:1986ys,Dolnik:1989qf,Eiswirth:1988ve,Glass:1984ly,Glass:2001eu,He:1989bh}.

A much richer behavior repertoire can be observed at medium-to-high levels of coupling strength. In general, stronger coupling leads to the deformation/skewness of the Arnol'd tongues, as well as to the intensity dependence of the phase shift \cite{Pikovsky:2001fk, Gonzalez:2008vn}. This in turn can lead to complex behaviors like period-doubling bifurcations, torus bifurcations, codimension-2 bifurcations, global bifurcations, bistability, cusps, and chaotic dynamics. Even though these behaviors have been extensively studied from a theoretical perspective, most of them still need to be described experimentally \cite{Gonzalez:2008vn}. Having this in mind, we decided to design and build an electronic master-slave oscillatory system (based in the 555 timer IC operating in the astable mode), and employed it to investigate synchronization in a wide range of values of forcing frequency and coupling strength. We also developed a mathematical model from the experimentally determined phase transition curve, and used it to increase our understanding of the synchronization phenomenon in this system. 

Our circuit design is inspired on closely related experimental and theoretical work
that has been carried out on forced astable multivibrators, some of which has even used the same
555 timer chip employed here \cite{Allen1983, Allen1983a, Tang1983}. In particular, we followed the works of \cite{Guisset2007, RamirezAvila2003, RamirezAvila2007, RamirezAvila2011, RamirezAvila2010, Rubido:2011rt, Rubido:2009kx, Rubido2011}, in which oscillators based upon the 555 IC are coupled through light. In these articles, the system behavior has been extensively studied from the point of view of phase-response curves, phase dynamics, noise influence, and Arnol'd tongues; and not only for the master-slave situation, but also for mutual-interaction. However, to the best of our knowledge, the authors have not studied synchronization in such a system in the strong coupling limit, and the present work is aimed at filling this void. To facilitate changing the coupling strength in a wide range of values, we decided to use electrical, instead of light, coupling. For the time being, we limit ourselves to the master-slave configuration, and leave other more general situations for future work. Our results suggest that the slave oscillator dynamics can be modeled by means of a circle map in which the phase transition curve has a horizontal segment, a feature that happens to be very important to understand its behavior. From this perspective, the present manuscript also draws from \cite{Belair:1986mz, Arnold1991, Gurney1992} who have extensively studied similar models.

\section{Circuit Design}
\label{circuit}

We are interested in studying the response of an electronic oscillator, built with a 555 timer IC, when it is periodically perturbed with short stimuli. The 555 timer is an integrated circuit with eight pins that are connected and/or have the functions described as follows:
\begin{itemize}
  \item Pin 1 is connected to the ground reference, or low level voltage ($0 \, \text{V}$).
  
  \item Pin 2 is an input pin called the trigger. The output pin goes high when this input falls below 1/2 of the control voltage, which is typically 2/3 of the positive supply voltage, Vs.
  
  \item Pin 3 is the 555 IC output pin. Its voltage can be equal to either Vs or $0 \, \text{V}$, depending on the state of the trigger, reset, and threshold pins.
  
  \item Pin 4 is the reset pin. Voltage at this input needs to be higher than about $1 \, \text{V}$ in order for Pins 3 and 7 to be able to switch between their available states.
  
  \item Pin 5 is an input pin that provides access to the control voltage. If this pin is unconnected, the control voltage takes its default value: 2/3 Vs. Usually this function is not required and the control input is often left unconnected. If electrical noise is likely to be a problem a 0.01µF capacitor can be connected between the control input and $0 \, \text{V}$ to provide some protection.
  
  \item Pin 6 is an input pin known as the threshold pin. Voltage at the output pin goes low when the voltage at this pin is greater than the control voltage (2/3 Vs if pin 5 unconnected).
  
  \item Pin 7 is called the discharge pin. It is an open collector output which may discharge a capacitor in phase with the output pin.
  
  \item Pin 8 is the power input pin. It must be connected to the positive supply voltage (Vs), which provides the power necessary for the IC to operate. 
\end{itemize}

\begin{figure}[ht]
\begin{center}
\includegraphics[width=2in]{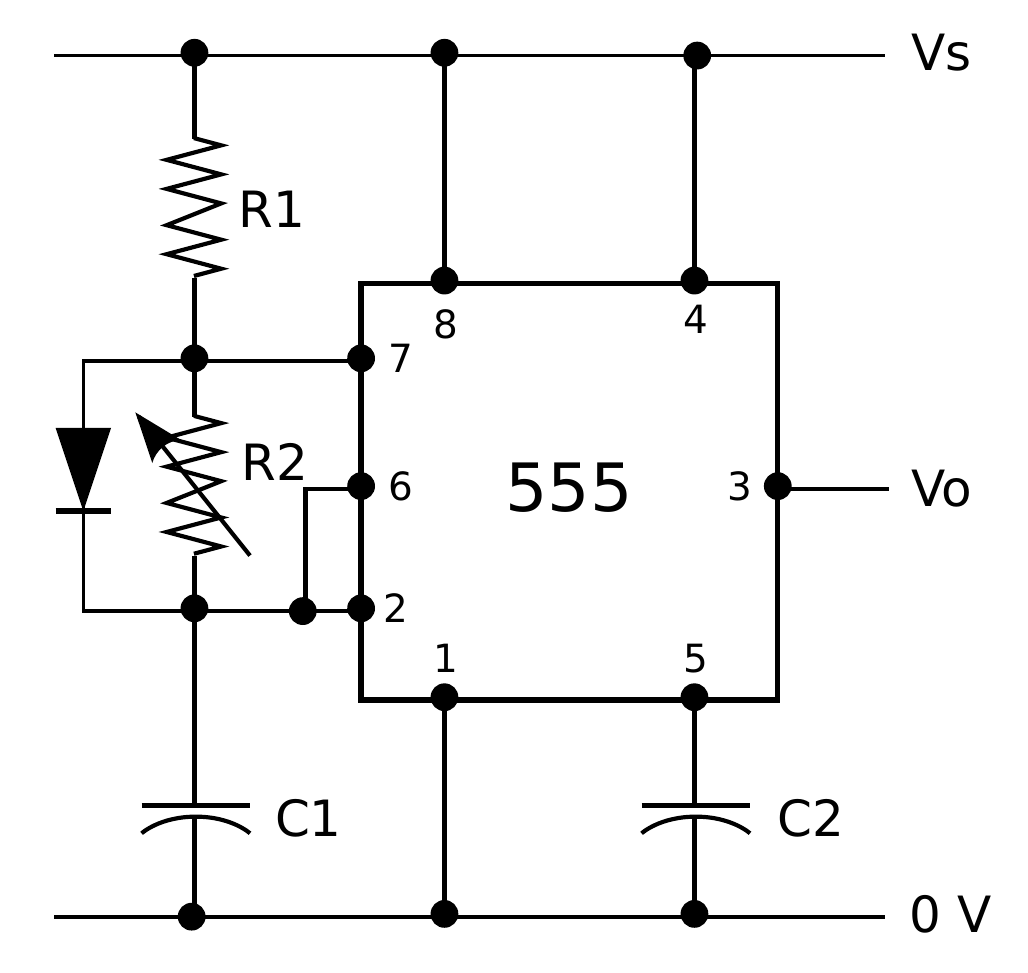}
\end{center}
\caption{Electronic oscillator built with a 555 timer IC.}
\label{Fig1}
\end{figure}

We employed the oscillator design illustrated in Fig. \ref{Fig1}. In accordance to the previous explanation of the 555-IC performance, the circuit oscillatory behavior can be explained as follows. Assume that capacitor C1 is initially discharged and that the voltage at pin 3 is high. Then, the capacitor starts charging through resistor R1 and the diode. When the voltage in the upper end of C1 exceeds 2/3 Vs, the voltage at pin 3 shifts to low and pin 7 gets connected to 0 V. This makes the capacitor discharge through resistor R2, until the voltage at the C1 upper end goes below 1/3 Vs. At this point, the voltage at pin 3 changes to high, pin 7 gets disconnected from 0V, and the cycle starts all over again. Notice that the diode in parallel with R2 permits the capacitor to charge through RI and discharge through R2. Henceforth, by properly choosing the values of both resistors one can control the charging and discharging times, which are respectively determined by the products R1 C1 and R2 C1. In particular, if R1 = R2 we obtain symmetric cycles.

\begin{figure}[h]
\begin{center}
\includegraphics[width=4in]{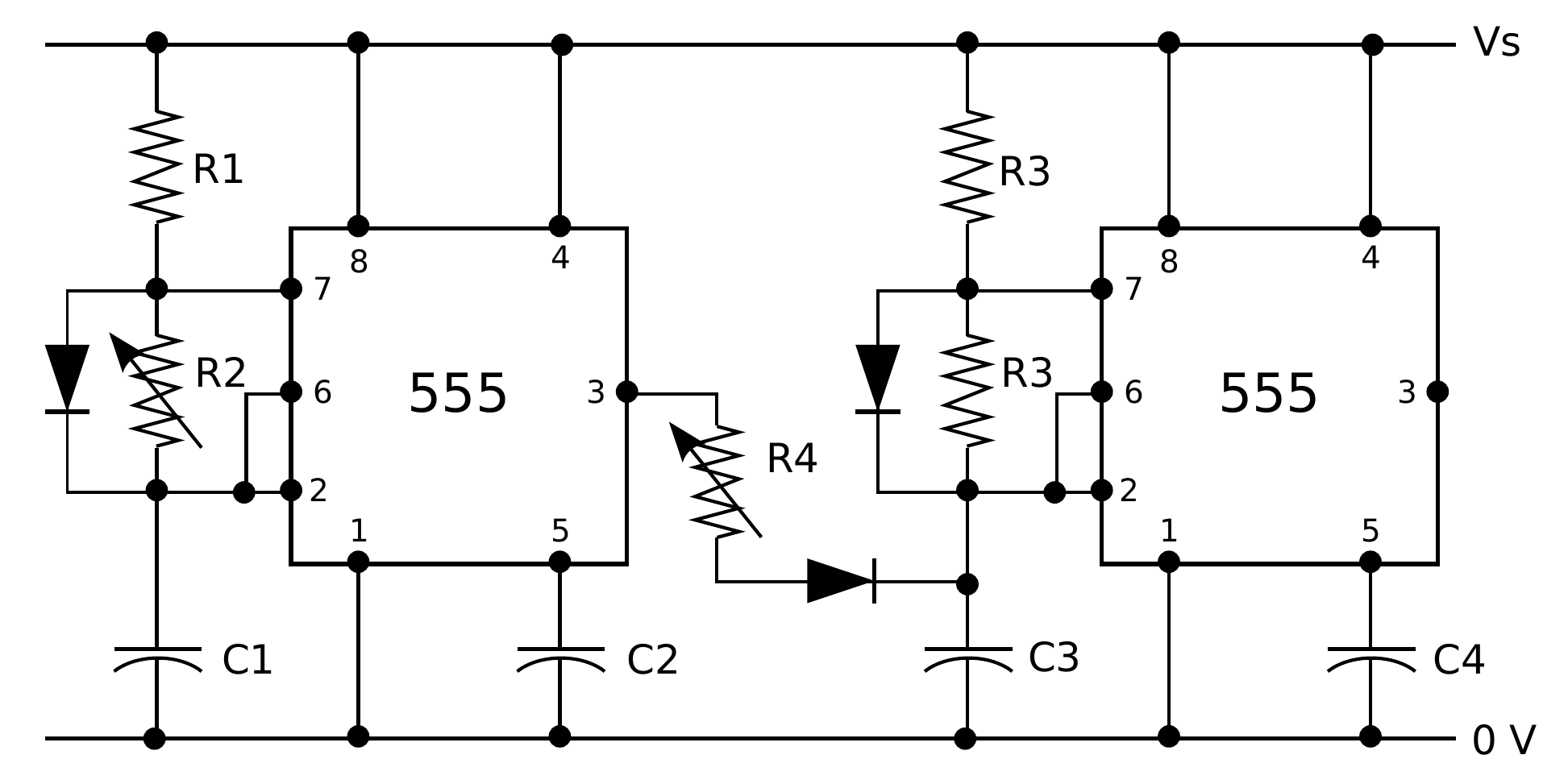}
\end{center}
\caption{Circuit design for two 555-timer-IC oscillators, one (to the left) periodically and unidirectionally perturbing the other (to the right).}
\label{Fig2}
\end{figure}

To study the response of a 555-timer electronic oscillator when it is periodically perturbed by short pulses we built the circuit represented in Fig. \ref{Fig2}. Basically, it consists of two oscillators like that in Fig. \ref{Fig1}, connected as a master-slave system. The one to the left corresponds to the master oscillator. It produces periodic pulses of fixed length (determined by resistance R1 and the capacitance C1), at a frequency determined by the product (R1 + R2) C1. The pulses from this oscillator are fed, through resistor R4 and a diode (that prevents the slave oscillator from influencing the master one), into capacitor C3 of the slave oscillator (the one to the right). When not perturbed, the slave oscillator produces symmetric cycles with a frequency determined by resistance R3 and capacitance C3. A pulse arriving from the master oscillator invariable increases the voltage of capacitor C3. The voltage increase depends on the values of R4 and C3. Then, if the pulse arrives while this capacitor is charging, it will shorten the charging time. On the contrary, if the pulse arrives during the discharging phase, the discharging time will get prolonged.

We wish to emphasize that the diode between resistor R4 and capacitor C3 ensures that electric current flows from the master oscillator into the slave one. In principle, this is sufficient to guarantee that the slave oscillator has no influence on the master one. Nonetheless, it could happen in practice that, when resistance R4 is very small, the current through it is so large that makes the voltage output at pin 3 (Vo) decrease. Recall that this pin voltage should be Vs during the pulse. Nevertheless, as we shall see in Section \ref{results}, our recordings indicate that such thing does not happen under the present conditions. On the other hand, there is also a voltage drop through the diodes when they are conducting. For the purpose of the present work, this effect can be regarded as equivalent to having extra voltage source of 0.6 V in series with R1 and R4. This in turn affects the current through such resistors. Despite that, our results are not affected because, as seen below, they rely upon direct measurements of the circuit dynamic variables, rather than on calculations depending on the resistance and capacitance values.

Summarizing, we designed and built the circuit in Fig. \ref{Fig2} to study the response of the slave oscillator when it is periodically stimulated (the pulse frequency can be controlled by varying resistance R2) with short pulses (pulse length being regulated by resistance R1) of varying intensity (the effect of the pulse on the slave oscillator is controlled by resistance R4).

The values of all the resistances and capacitances, as well as that of Vs, are tabulated in Table 1. With these parameters, the experimentally-measured duration of the pulses produced by the master oscillator is about 0.125 ms, while the frequency of the slave oscillator is about 180 Hz. That is, the slave oscillator period is more than forty times larger than the duration of the master oscillator pulses. We varied the value of resistance R2 so that the frequency of the forcing stimulus changed between about 0.87 and 1.73 times the frequency of the slave oscillator. Resistance R4 was varied in the range from 78.125 $\Omega$ to 10 k$\Omega$. Finally we employed the 1N4148 signal diode, which can conduct up to 0.3 A and has a reverse breakdown voltage of 100 V.

\begin{table}
\tbl{Parameter values for the circuit in Fig. \ref{Fig2}}
{\begin{tabular}{cc|cc}\\[-2pt]
\hline
Parameter & Value & Parameter & Value \\
\hline
R1 & 180 $\Omega$ & C1 & 1 $\mu$F \\
R2 & 0-10 k$\Omega$ & C2 & 0.1 $\mu$F \\
R3 & 4 k$\Omega$ & C3 & 1 $\mu$F \\
R4 & 0-10 k$\Omega$ & C4 & 0.1 $\mu$F \\
Vs & 5 V \\
\hline
\end{tabular}}
\end{table}

\section{Data Acquisition and Experimental Results}
\label{results}

Data, in the form of voltage time series measured at pin 3 of the master-oscillator 555 timer IC and at the upper end of capacitor C3 in the slave oscillator (see Fig \ref{Fig2}), were acquired by means of a \texttt{BitScope} BS10 (manufactured by BitScope\footnote{http://www.bitscope.com}), controlled from \texttt{Python 2.7} via the public library \texttt{BitScope Library} 2.0. We employed a sampling frequency of 80 kHz, and obtained 12,000 data points each time we recorded. This guaranteed the acquisition of more than 25 cycles of the slave oscillator, with enough detail to capture the beginning and the end of the forcing pulses.

\begin{figure}[ht]
\begin{center}
\includegraphics[width=5in]{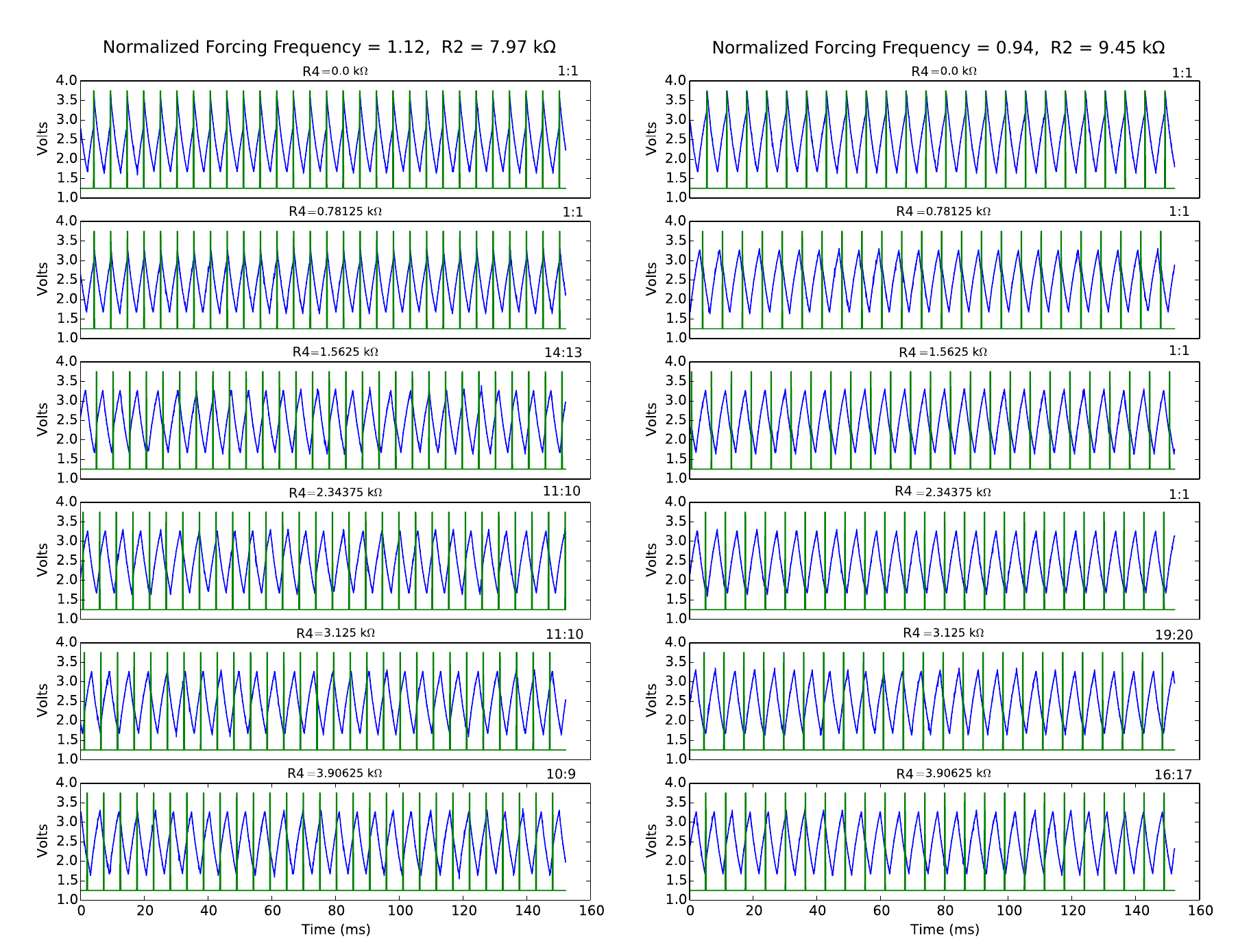}
\end{center}
\caption{Plots of the voltage measured at pin 3 of the 555 timer IC corresponding to the master oscillator (green lines), and at the positive end of capacitor C3 in the slave oscillator (blue lines), for two different normalized forcing frequency values (resistance R2), and for various values of resistance R4. The rotation number corresponding to the phase-locking rhythm achieved in each case is shown.}
\label{Fig3}
\end{figure}

We carried out the measurements described in the previous paragraph for several combinations of resistances R2 and R4 in the range $[0 \, \Omega, 10 \, \text{k}\Omega]$. Recall that the frequency of the forcing pulse can be changed by modifying resistance R2, while resistance R4 determines how strongly the forcing pulse affects the slave oscillator. The results for two specific values of R2 and various R4 values are illustrated in Fig. \ref{Fig3}. Notice that, since the frequency of the uncoupled slave oscillator remains unchanged in all of our experiments, it is sufficient to indicate the value of the master oscillator frequency normalized to that of the slave oscillator (here and thereafter we shall refer to it as the normalized forcing frequency) to know both of them.

\begin{figure}[ht]
\begin{center}
\includegraphics[width=3.25in]{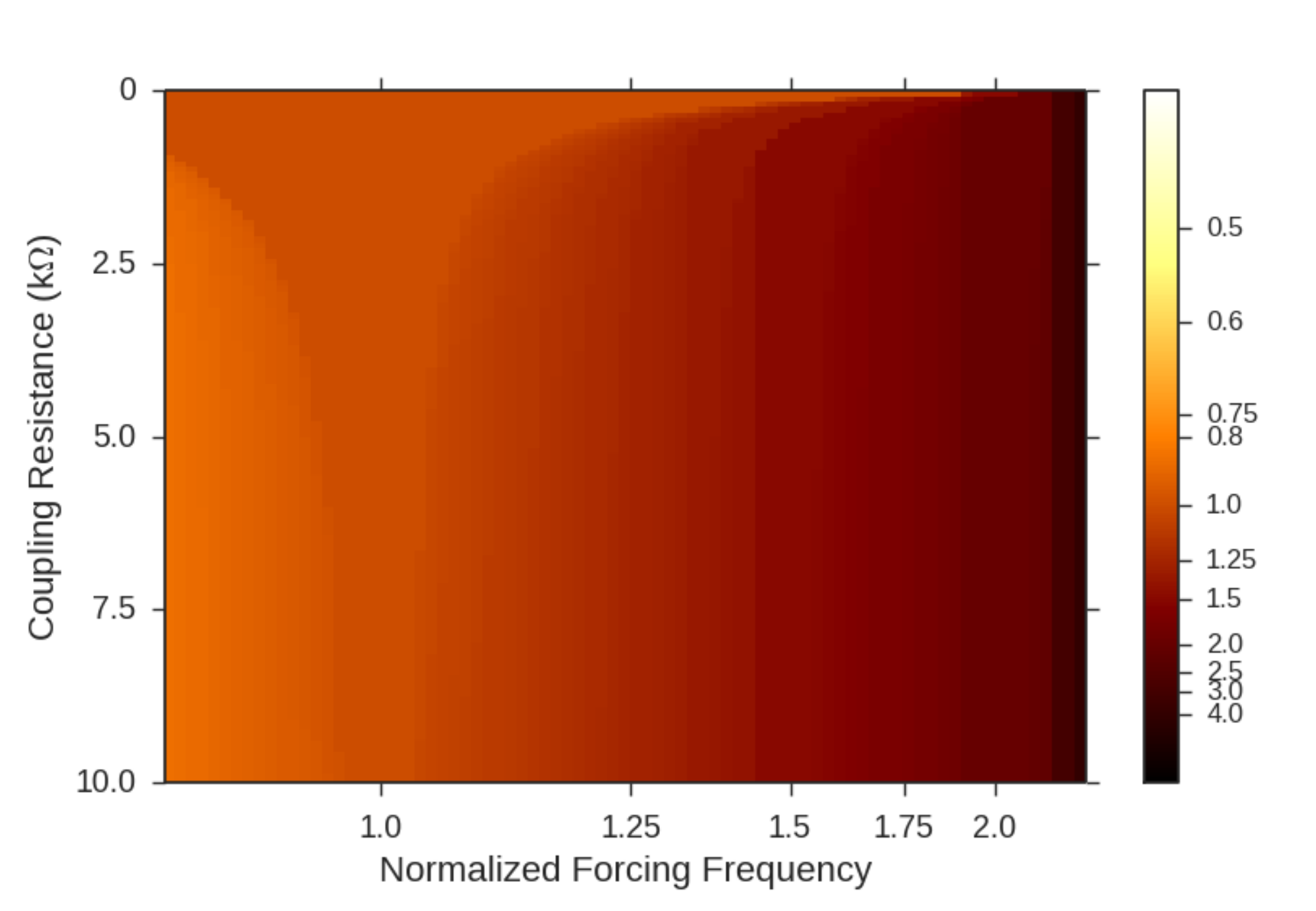}
\end{center}
\caption{Plot the rotation number corresponding to the phase-locking rhythms achieved by the master-slave oscillator system vs. the normalized forcing frequency and the value of the coupling resistance R4. The color code indicating the rotation number is shown in the right-hand-side bar.}
\label{Fig4}
\end{figure}

Observe that, when the value of resistance R4 is very low, the master oscillator is capable of forcing the slave oscillator to cycle at its same frequency. However, as the value of R4 increases, the phase-locking rhythm becomes more intricate. In general, we can observe complex cycles in which the system behavior repeats after $m$ pulses of the master oscillator and $n$ pulses of the slave one. This situation is called an $m:n$ phase-locking rhythm. 

\begin{figure}[ht]
\begin{center}
\includegraphics[width=3in]{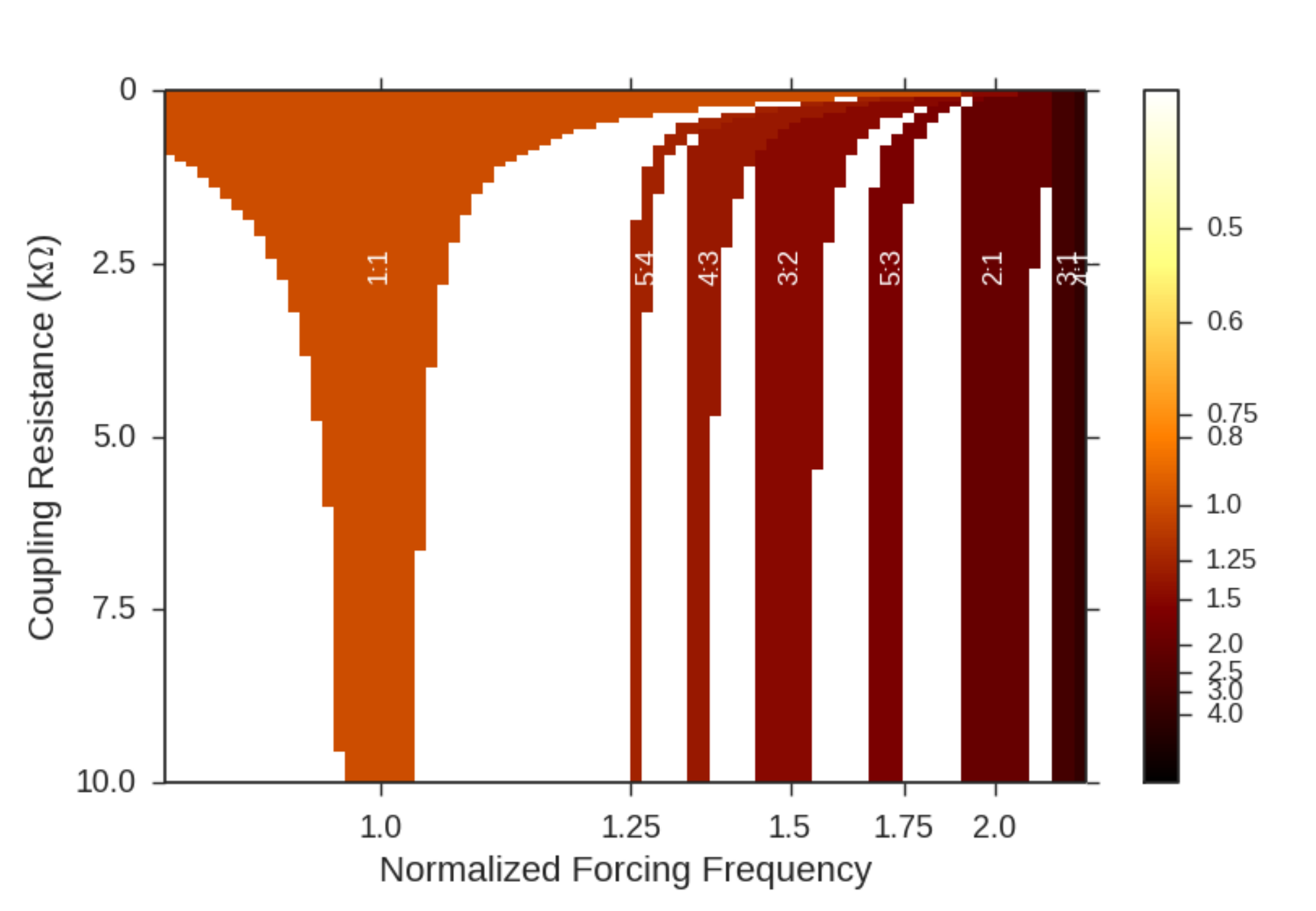}
\end{center}
\caption{Plot of a few Arnol'd tongues extracted from the data in Fig. \ref{Fig4}. The corresponding phase-locking rhythms are indicated within the Arnol'd tongues.}
\label{Fig5}
\end{figure}

We repeated the measurements illustrated in Fig. \ref{Fig3} for several values of resistances R2 and R4, and determined in each case the normalized forcing frequency and the achieved phase-locking rhythm. The obtained results are summarized in Fig. \ref{Fig4} where the rotation number is plotted---by means of a color code---vs. the normalized forcing frequency and the value of the coupling resistance R4. Here, we define the rotation number as the average number of cycles of the master oscillator divided by the average number of cycles of the slave oscillator (notice that the inverse definition is used in many other papers). From this definition, the rotation number corresponding to the phase-locking rhythm $m:n$ is the ratio $m/n$. We can appreciate that Fig. \ref{Fig4} contains the full set of Arnol'd tongues for the 555-timer oscillator, or at least as many as the experimental resolution permitted to elucidate. Although Fig. \ref{Fig4} nicely illustrates how bifurcations affecting the rotation number take place as we move in the parameter space, it is somehow difficult to identify specific tongues due to small color gradients. It is also possible that that different Arnol'd tongues are represented with the same color because all phase locking rhythms of the form $i \cdot m : i \cdot n$ (with $i$ integer) have the same rotation number. To improve our understanding of the Arnol'd tongues behavior we isolated the tongues up to order 5 and show them in Fig. \ref{Fig5}. We can see there that all the tongues to the left of that corresponding to the $2:1$ phase locking rhythm move up and to the right to presumably terminate in a degenerate bifurcation point \cite{Belair:1986mz}. The tongues to the right of the $2:1$ one behave differently. We only have one clearly-visible tongue: $3:1$. Interestingly, in the high-coupling-resistance (weak coupling) limit, it does not end at the normalized forcing frequency that is equal to the corresponding rotation numbers ($3$ in this case), as the other tongues do and as one would expect from the theory \cite{Boyland:1986ys}. We believe this is due to the fact that at such high frequencies of the master oscillator, the short pulse approximation is no longer valid; i.e. the time period between pulses is comparable to the pulse duration.

\section{Mathematical Modeling}
\label{model}

\begin{figure}[ht]
\begin{center}
\includegraphics[width=3in]{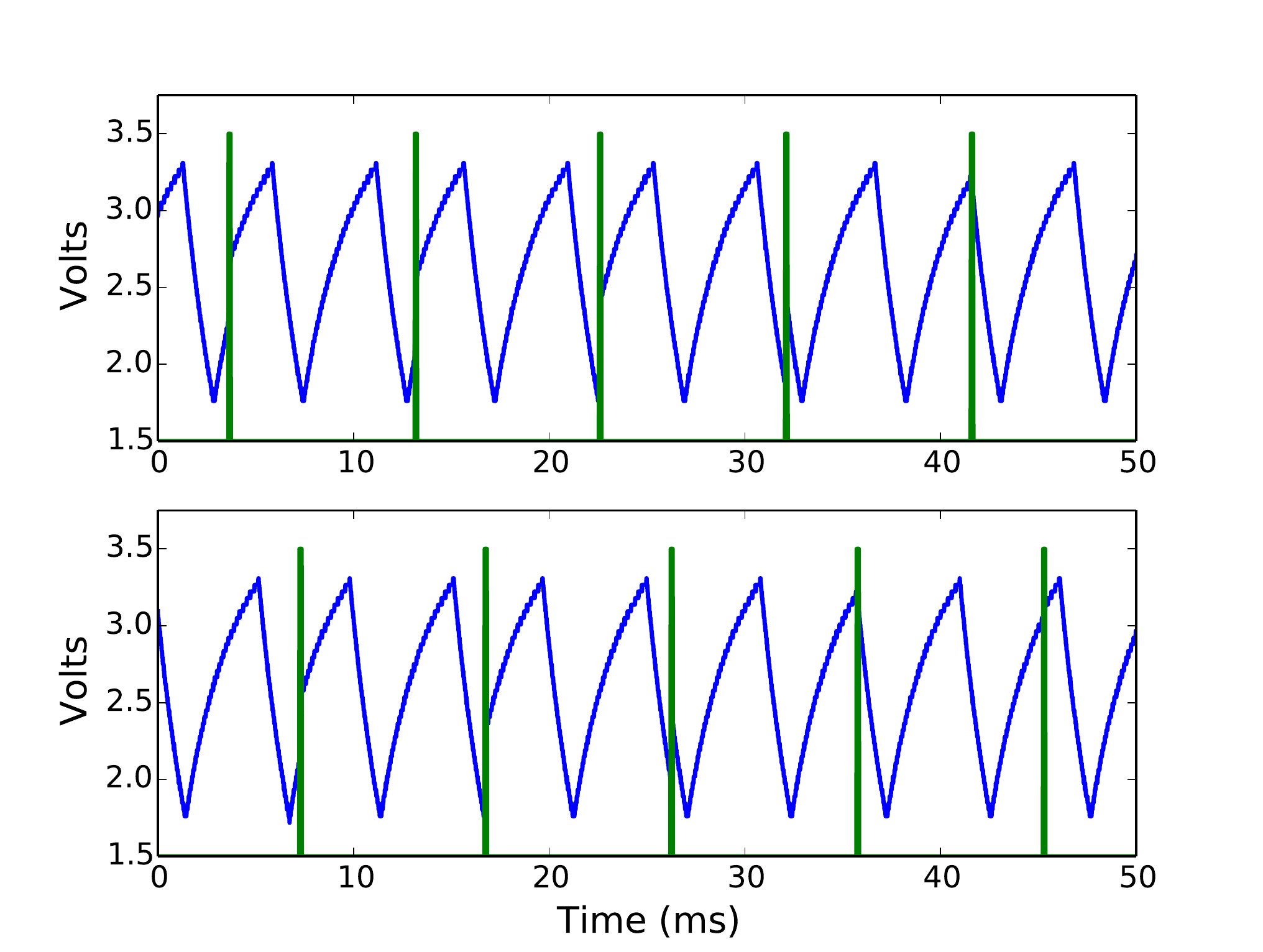}
\end{center}
\caption{Two different recordings, made under identical conditions, of voltage at pin 3 (green lines) of the master oscillator and at the upper end of capacitor C3 (blue lines) in the slave oscillator, when the master oscillator frequency is almost twice as large as that of the slave oscillator. Indeed, the normalized forcing frequency is 1.82. The coupling resistance R4 was set to $300 \, \Omega$.}
\label{Fig5-5}
\end{figure}

In Fig. \ref{Fig5-5} we show recordings of pulses produced by the master oscillator and of the corresponding slave-oscillator response. The master oscillator frequency was set to be almost twice as large as that of the slave oscillator, and the coupling resistance R4 was set to $300 \, \Omega$. We can appreciate in the plots of Fig. \ref{Fig5-5} that the the pulse effect depends on the slave-oscillator phase at which it arrives. Nonetheless, in all cases, only one cycle in the slave oscillator is affected, while all subsequent cycles remain unchanged. This allows us to mathematically model the slave-oscillator dynamics as a circle map.

From the discussion in the previous paragraph, we can assume that the effect of a single pulse is to instantaneously reset the slave oscillator phase. Let $\phi$ denote the old phase of the slave oscillation immediately preceding the perturbation, and $\theta$ be the new phase immediately following the perturbation. Hence, the slave oscillator response to a pulse of intensity $h$ can be summarized in a function of the form:
\[
\theta = g(\phi, h),
\]
which is usually called the phase transition curve (PTC) \cite{Guevara:1982rr}.

\begin{figure}[ht]
\begin{center}
\includegraphics[width=3in]{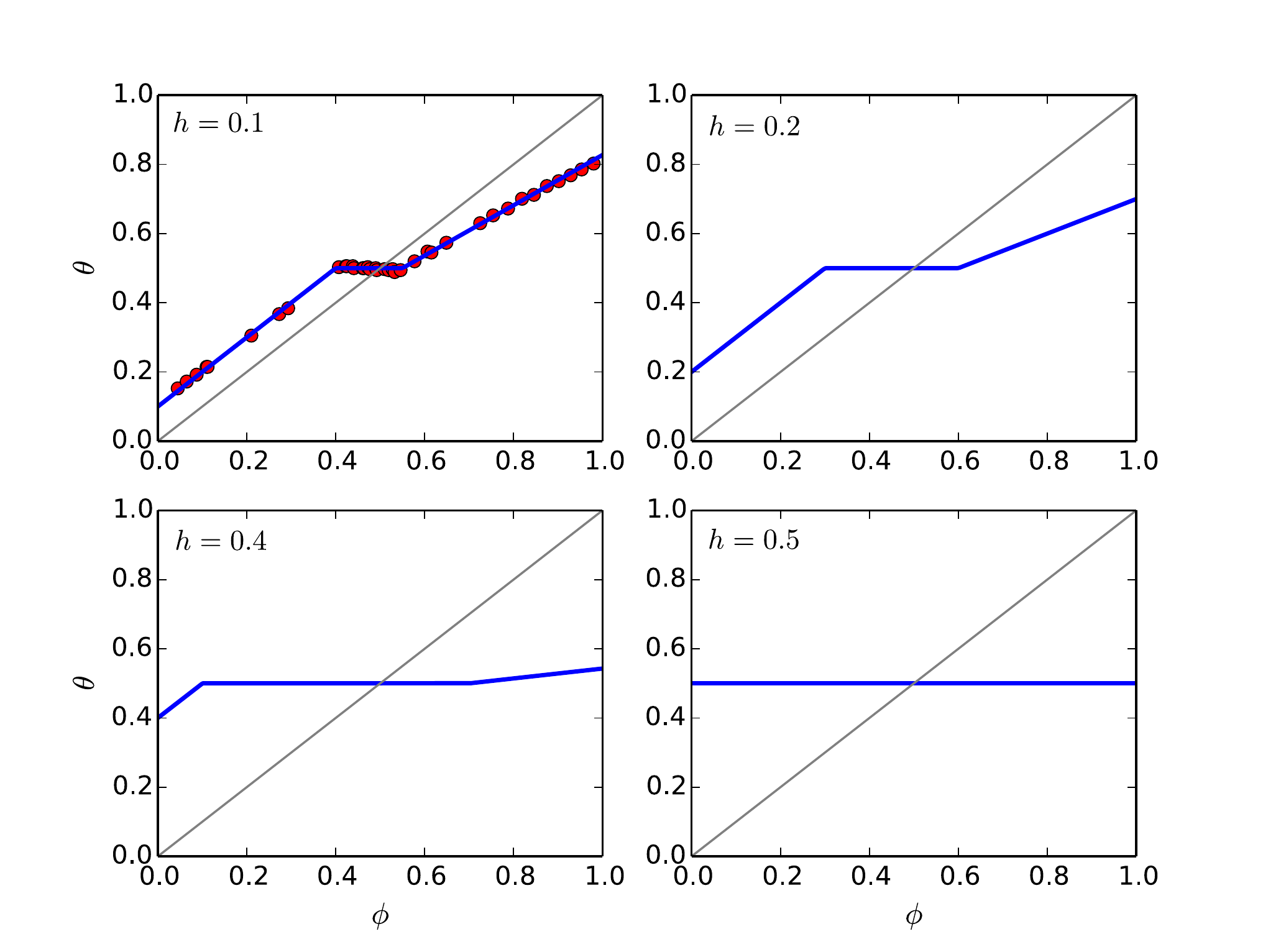}
\end{center}
\caption{In the upper-left panel we show the experimentally measured phase transition curve (PTC)---circles---obtained with R4 = 9.14 k$\Omega$; together with the corresponding piece-wise linear fitting function $g(\phi, h)$---solid line---given by Eq. (\ref{eq01}) with $h = 0.1$. Variables $\phi$ and $\theta$ respectively denote the slave-oscillator phase immediately preceding and following the perturbation. In the rest of the panels we plot function $g(\phi, h)$ with increasing  values of $h$.}
\label{Fig6}
\end{figure}

To find the 555-timer oscillator PTC, we analyzed the recordings corresponding to the coupling resistance R4 = 9.14 k$\Omega$ and to several values of resistance R2, so as to change the frequency of the master oscillator. In each case, we measured the phase at which the pulse arrives, as well as how much the corresponding slave-oscillator cycle was shortened or elongated---here and thereafter we considered that the cycle starts ($\phi = 0$) at the beginning of the charging phase of capacitor C3. If the cycle is shortened (elongated), this means that the oscillator phase was advanced (retarded) due to the pulse. By repeating these measurements at different phases, we were able to experimentally determine several points of the PTC and plotted them in Fig. \ref{Fig6}. Notice that the experimental PTC points corresponding to the selected coupling-resistance value can be approximated by a piece-wise linear function of the form:
\begin{equation}
g(\phi,h') = \left\{
\begin{array}{ll}
\phi + h', & {\rm if} \quad 0 \leq \phi \leq \frac{1}{2} - h', \\
\frac{1}{2}, & {\rm if} \quad \frac{1}{2} - h' < \phi < \frac{1}{2} + \frac{h'}{2}, \\
h' + \frac{1/2-h'}{1/2+h'/2}\phi, & {\rm if} \quad \frac{1}{2} + \frac{h'}{2} <  \phi \leq 1,
\end{array}
\right.
\label{eq01}
\end{equation}
with $h' = 0.1$.

The functional form of Eq. \ref{eq01} suggests that parameter $h'$ can be interpreted as the pulse intensity $h$. To further investigate this assertion, note that the PTC has a flat region around $\phi = 0.5$; which can be explained as follows. According to our observations, the phase $\phi  = 0.5$ corresponds to the point at which capacitor C3 in the slave oscillator stops charging and starts discharging. Taking this into account, we can see that the PTC flat region corresponds to initial slave-oscillator phases ($\phi$) such that the pulse from the master oscillator makes capacitor C3 start discharging. That is, if this capacitor is charging, the pulse increases its charge up to the discharging threshold value; and if C3 is already discharging, the pulse charges it back to the beginning of the discharging stage. From these considerations, we expect that the PTC flat region widens as the coupling resistance R4 decreases; the reason being that, as R4 diminishes, the current through it increases, and so the charge increase in capacitor C3 due a single pulse enlarges. On the other hand, the PTC flat region also widens as parameter $h'$ increases---see Fig. 7. This, together with the fact that such parameter also corresponds to the value returned by function $g(\phi, h')$ when $\phi = 0$---see Eq. (1), supports the interpretation of $h'$ as the intensity of the perturbing pulse. Consequently, we assume that Eq. (1) is a good approximation to the PTC for all values of the coupling resistance (R4), and that R4 decrements can be modeled by concomitant $h'$ increments. In particular, we suppose that resistance R4 is inversely proportional to $h'$, which corresponds to the pulse intensity $h$.

Once the slave-oscillator phase transition curve has been determined, its discrete-time dynamics in response to periodic forcing are dictated by the following finite-difference equation \cite{Glass:1988wd,Kaplan:1995lq}:
\begin{equation}
\phi_{i+1} = {\rm mod}\left(\; [\; g(\phi_{i},h) \; + {\rm mod}(\tau, 1) \; ], \; 1 \right),
\label{eq02}
\end{equation}
in which ${\rm mod}$ denotes the modulus operation, $\phi_i$ represents the slave-oscillator phase at the time it receives the $i$ th pulse of intensity $h$, and $\tau$ is the time period between one forcing pulse and the next, normalized to the slave-oscillator period when not perturbed. The value of $\tau$ can be related to the component values in the circuit as follows:
\[
\tau \simeq \frac{(\text{R1} + \text{R2}) \, \text{C1}}{2 \, \text{R3} \, \text{C3}}.
\]

\begin{figure}[ht]
\begin{center}
\includegraphics[width=3in]{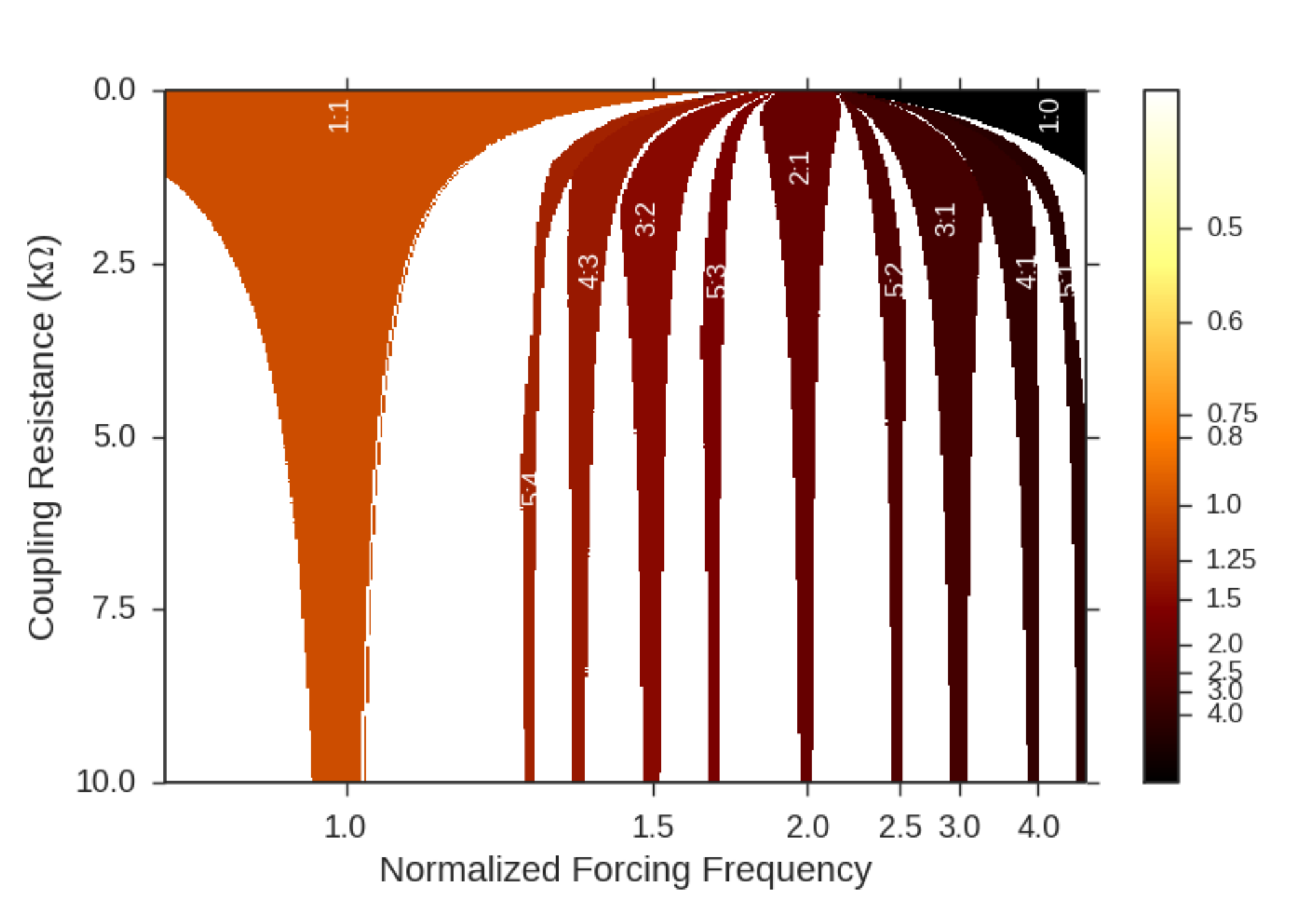}
\end{center}
\caption{Plots of Arnol'd tongues, up to order 5, computed from the model in Eq. (\ref{eq02}), in terms of the normalized forcing frequency and coupling resistance value (R4). In the corresponding calculations we assumed that parameter $h$ decreases proportionally to R4 increments, that $h=0.5$ when R4 = 0, and that $h=0.024$ when R4 = 10 k$\Omega$.}
\label{Fig7}
\end{figure}

To test the model feasibility, we computed the corresponding Arnol'd tongues up to order five. The results are shown in Fig. \ref{Fig7}. To carry out the respective calculations we took into account that parameter $h$  decreases inversely proportionally to resistance R4, and assumed that $h=0.5$ when R4 = 0, and that $h=0.024$ when R4 = 10 k$\Omega$. We further considered plentiful different points in the coupling-resistance vs. normalized-forcing-frequency parameter space, and computed the achieved phase-locking rhythm for each one of them. To find the phase locking rhythms we iterated Eq. (\ref{eq02}) three hundred times, and ignored the first two hundred iterations in order to guarantee that the system reaches a stationary behavior. If the resulting phase locking rhythm was of the form $m:n$, with $m=1,2,3,4,5$, we plotted it in the graph of Fig. \ref{Fig7}, otherwise we ignored it. We repeated each point computation twenty times---starting from randomly selected initial conditions---to test for multistability, but we found no sign of it given that each time we obtained the same result. This uniqueness of solution might be due to the fact that the middle branch of the model PTC has zero slope, so that there is no non-monotonicity in the PTC---see for instance type A vs. type B maps in \cite{Belair:1986mz}. This result might change should a more realistic model be used.
\begin{figure}[ht]
\begin{center}
\includegraphics[width=3in]{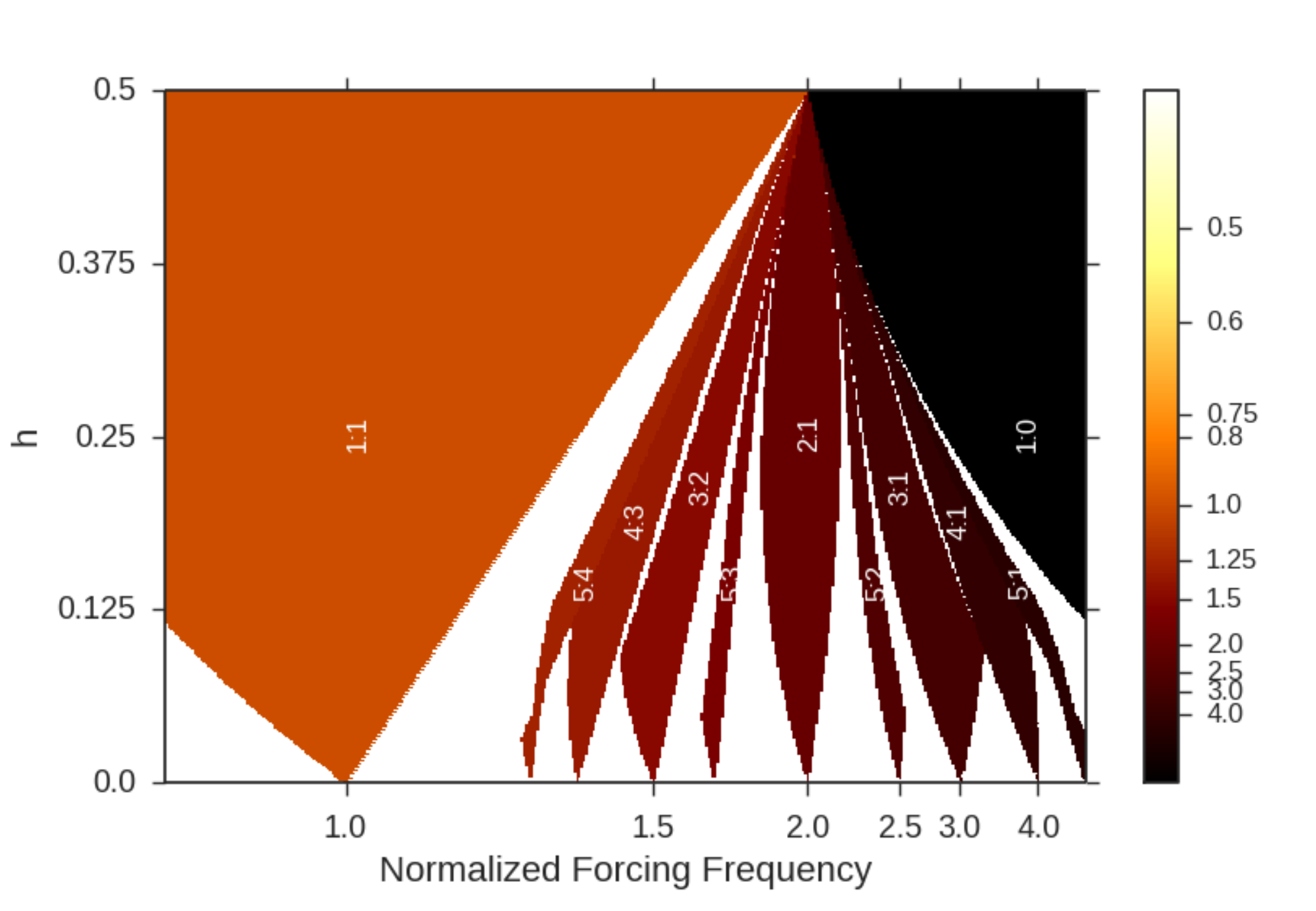}
\end{center}
\caption{Plots of Arnol'd tongues, up to order 5, computed from the model in Eq. (\ref{eq02}), in terms of the normalized forcing frequency and of parameter $h$.}
\label{Fig8}
\end{figure}

Observe that for values of the normalized forcing frequency (NFF) smaller than 2. the Arnol'd tongues in Figs. \ref{Fig5} and \ref{Fig7} are quite similar. Regarding the differences, we can see in Fig. \ref{Fig7} that for every tongue of the form $m:n$ to the left of NFF = 2 there is a symmetrically placed tongue of the form $m:n'$ to the right. Furthermore, all tongues rise and curve towards NFF = 2, where they apparently converge. To better appreciate this behavior, we numerically computed the Arnol'd tongues again, but now as a function of the normalized forcing frequency and of parameter $h$. The results are shown in Figure \ref{Fig8}---recall that $h$ changes are assumed to be inversely proportional to R4. There, we can observe how all tongues converge to NFF = 2 when $h=0.5$, indicating that this is a codimension-2 degenerate bifurcation point.

The existence of this degenerate bifurcation point can be understood by noticing in Fig. \ref{Fig6} that the PTC becomes a flat horizontal line when $h=0.5$. This means that, in the limit of very strong coupling, any pulse will drive the slave oscillator to $\phi = 1/2$---regardless of the phase at which it arrives---and the slave-oscillator capacitor shall start discharging immediately after each pulse. All this means that, if the master-oscillator frequency is larger than twice that of the slave oscillator, the pulses will always arrive while the slave capacitor is discharging, immediately taking the capacitor back to the beginning of the discharging phase. In consequence, the pulses won't let the slave oscillator complete any cycle. This explains the 1:0 phase locking rhythm. Contrarily, if the master oscillator frequency is smaller than twice that of the slave oscillator, the pulses will arrive when the slave capacitor is charging; a given pulse will completely charge the capacitor; and it will start discharging immediately. The next pulse will arrive when the capacitor is charging, and the cycle shall be repeated. That is, every pulse makes the oscillator complete a cycle, and this explains the 1:1 phase-locking rhythm seen in the upper-right corners of Figs. \ref{Fig7} and \ref{Fig8}.

\section{Strong-coupling limit}
\label{scl}

In order to experimentally corroborate the model prediction that a codimension-2 bifurcation point exists in the very-strong coupling limit, we modified our experimental setup as follows:

\begin{figure}[ht]
\begin{center}
\includegraphics[width=3in]{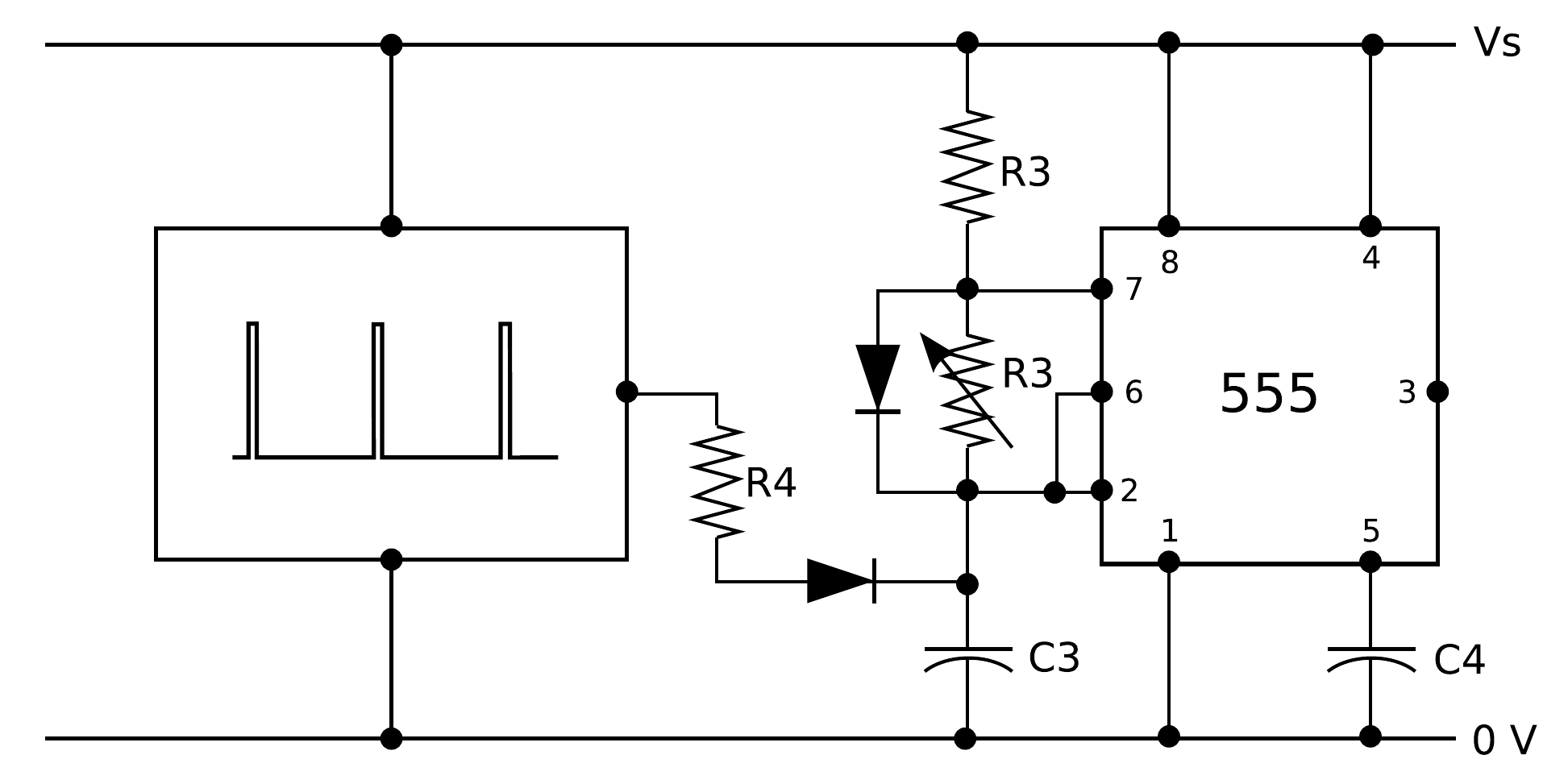}
\end{center}
\caption{Experimental setup designed to investigate the synchronization of the master-slave system here introduced in the strong coupling limit.}
\label{Fig10}
\end{figure}

\begin{itemize}
  \item The model predicts that, in the limit of very strong coupling, a sharp transition occurs between the $1:1$ and the $1:0$ phase-locking rhythms, as the normalized forcing frequency moves from $1$ to $\infty$. However, it is not straightforward to experimentally determine when the slave oscillator shows no complete oscillations, and hence when the $1:0$ phase-locking rhythm is achieved. To solve this problem we took advantage of the fact that, due to the modulus function in Eq. (\ref{eq02}), the same behavior should be observed between the $1:2$ and the $1:1$ phase-locking rhythms. Thus, in the modified experimental protocol, we chose the normalized forcing frequency to take values around the interval $[0.5,1]$, to experimentally determine this second bifurcation point.
  
  \item To improve our control over the master oscillator, we substituted it by a microcontroller programed to produce short pulses, as schematically represented in Fig. \ref{Fig10}. By changing the microcontroller program parameters we can modify pulse duration and pulse frequency. In particular, we considered pulse durations in the interval $[1,4] \, \hbox{ms}$.
  
  \item To guarantee that the short-pulse approximation is always satisfied, we chose the following values for resistance R3 and capacitance C3: $\hbox{R3} = 2.2 \, \hbox{k}\Omega$ and $\hbox{C3} = 10.0 \, \mu \hbox{F}$. With this component values, the slave oscillator frequency is about $17 \, \hbox{Hz}$.
  
  \item To have a finer control over the coupling strength, we fixed the coupling resistance value to $\text{R4} = 220 \, \Omega$, and varied instead the duration of the pulses produced by the microcontroller. The rationale for doing this was as follows: a longer pulse duration means that current flows into the slave-oscillator capacitor for a longer time, and hence that more charge accumulates in it due to the pulse. This in turn implies that a larger voltage increase is  achieved.
\end{itemize}

\begin{figure}[ht]
\begin{center}
\includegraphics[width=3in]{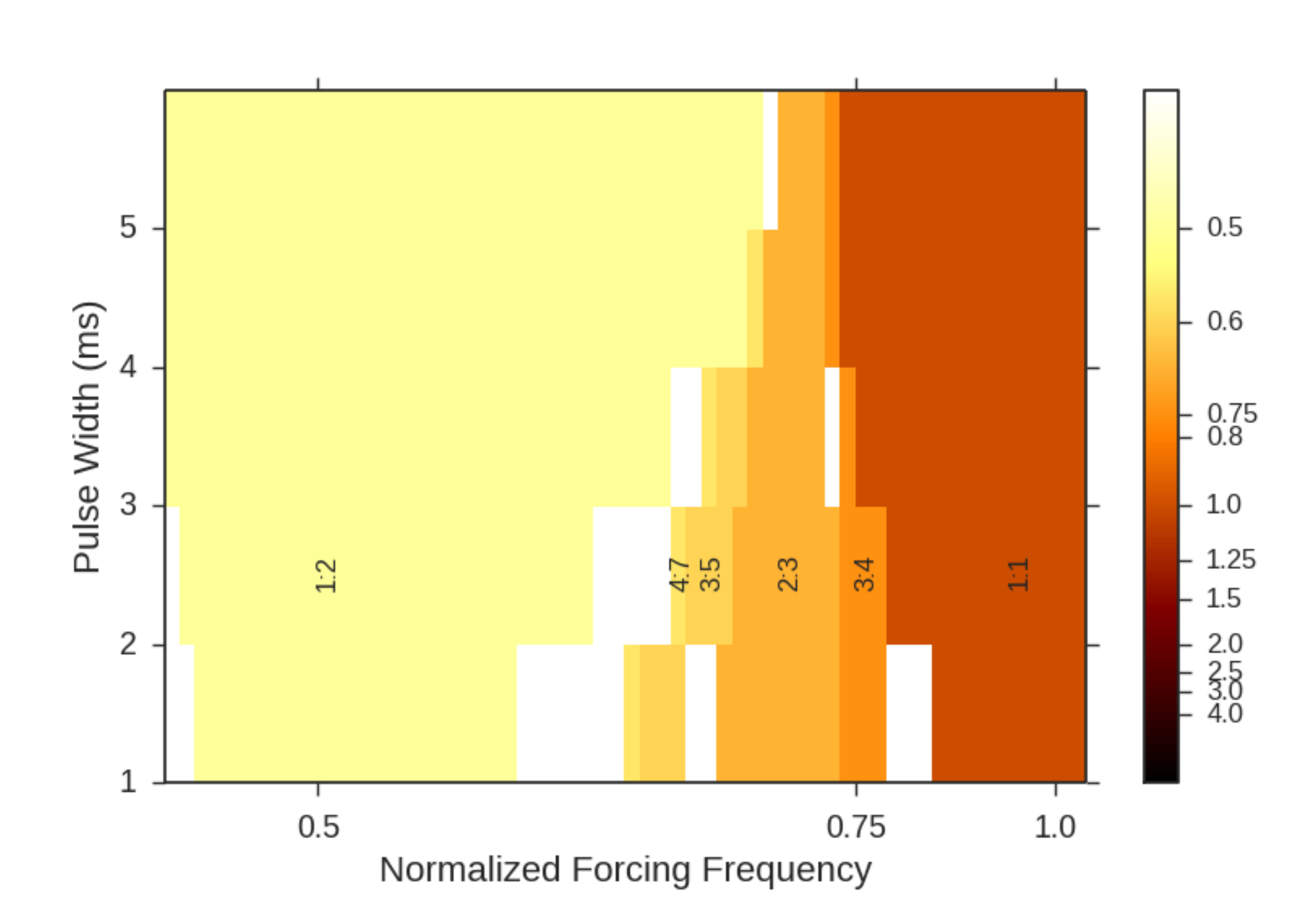}
\end{center}
\caption{Plots of Arnol'd tongues, up to order 4, experimentally determined from the experimental setup represented in Fig. \ref{Fig10}. Recall that this setup was explicitly designed to inspect the system behavior in the strong coupling limit.}
\label{Fig11}
\end{figure}

After adjusting the experimental protocol, we repeated the procedure previously employed to determine the system Arnol'd tongues up to order 4, with the only difference that the data acquisition card was set to 10,000 samples per second. The results are shown in Fig. \ref{Fig11}. Notice how the experimental results agree with the model predictions (see Fig. \ref{Fig8}). Namely, the $1:2$ and the $1:1$ regions expand as the coupling strengthens (squeezing all the tongues in between), and they tend to converge in a codimension-2 bifurcation point at the very-strong coupling limit.

\section{Concluding remarks}
\label{conclu}

We designed and built a master-slave electronic oscillatory system based on the 555-timer oscillator---see Fig. \ref{Fig2}, and studied the synchronization of this system for several values of the master slave frequency and of the coupling resistance (the uncoupled slave oscillator frequency was kept constant throughout the experiments). Our results demonstrated the existence of phase-locking regions (also known as Arnol'd tongues) in the parameter space, all of which apparently converge at NFF = 2 and R4 = 0. This would imply that the the system phase locking rhythm changes, via period-adding bifurcations, as we move in the parameter space, with the exception of the NFF = 2 and R4 = 0 which is a degenerate bifurcation point. These results were confirmed by a circle map built to model the response of a 555-timer oscillator to periodic perturbations.

The existence of degenerate codimension 2 bifurcation points into which the Arnol'd tongues collapse can be understood in terms of the circle-map phase transition curve (PTC), which has a horizontal middle segment. According to Eq. (\ref{eq01}), this middle segment widens as the coupling between the master and the slave oscillators strengthens until, in the very strong coupling limit, the whole PTC is a horizontal line. As previously discussed, this behavior of the PTC explains the collapse of Arnol'd tongues. This result is in complete agreement with the findings of previous works in which similar circle maps have been studied \cite{Allen1983, Allen1983a, Belair:1986mz, Arnold1991, Gurney1992}. Although the appearance of degenerate codimension 2 bifurcation points had been theoretically predicted and some experimental evidence had been collected, to the best of our knowledge, this is the first time that it is experimentally demonstrated by thoroughly exploring the normalized forcing frequency-coupling strength parameter space. Interestingly, modeling the system dynamics via a circle map with a piece-wise linear PTC is only a rough approximation. A more detailed model would consist of ordinary differential equation that take into account the dynamics of all the circuit components. However, such a model would not lead to a piece-wise linear PTC with a flat region, and so it would not predict a degenerate codimension point. Taking this into consideration, we can conclude that we have experimentally observed this bifurcation either because the exploration has not been made on a fine enough scale to distinguish the individual termination points of the Arnol'd tongues at high forcing intensity on the boundaries of other tongues, or because such fine-scale structure is not determinable in this circuit, given the amount of experimental noise that is inevitably present.

Finally, we wish to emphasize the didactic value of the system here analyzed. Given the facility to build it, the non expensiveness of its components, and the general availability of the equipment necessary to carry out the corresponding measurements, it can be employed as a laboratory exercise to exemplify the usage of nonlinear dynamics to understand the phenomenon of synchronization.

\section*{Acknowledgements}

\noindent
The author deeply thanks Profs. Michael C. Mackey and Michael Guevara for their valuable advice, McGill University for its hospitality while carrying out this project, and D. Orozco-Gómez and E. Sosa-Hernández for critically reading the manuscript. The author is also grateful to the anonymous reviewers whose comments greatly helped to improve the manuscript.

\bibliography{555Synch}

\end{document}